\documentclass[11pt]{article}

\usepackage[preprint]{acl}

\usepackage[T1]{fontenc}
\usepackage[utf8]{inputenc}
\usepackage{times}
\usepackage{latexsym}
\usepackage{microtype}
\usepackage{inconsolata}

\usepackage{xcolor}
\usepackage{graphicx}
\usepackage{booktabs}
\usepackage{multirow}
\usepackage{siunitx}
\usepackage{colortbl}
\usepackage{pifont}

\usepackage{amsmath,amssymb,amsthm}

\usepackage{algorithm}
\usepackage{algpseudocode}

\usepackage{enumitem}
\usepackage[most]{tcolorbox}

\tcbset{
  aibox/.style={
    width=400.18663pt,
    top=10pt,
    colback=white,
    colframe=black,
    colbacktitle=black,
    enhanced,
    center,
    attach boxed title to top left={yshift=-0.1in,xshift=0.15in},
    boxed title style={boxrule=0pt,colframe=white,},
  }
}
\newtcolorbox{AIbox}[2][]{aibox,title=#2,#1}

\title{EvA: An Evidence-First Audio Understanding Paradigm for LALMs}

\author{Xinyuan Xie$^{1,2}$\thanks{Equal Contribution.}~, ~Shunian Chen$^{1}$\footnotemark[1]~, ~Zhiheng Liu$^{1}$, ~Yuhao Zhang$^{1}$ \\ \textbf{Zhiqiang Lv$^{2}$}, ~\textbf{Liyin Liang$^{2}$}, ~\textbf{Benyou Wang$^{1}$}\thanks{Corresponding author.}\\
The Chinese University of Hong Kong, Shenzhen$^{1}$,~Didi Chuxing$^{2}$\\
\texttt{wangbenyou@cuhk.edu.cn} \\
}

\begin{document}
\maketitle
\begin{abstract}
Large Audio Language Models (LALMs) still struggle in complex acoustic scenes because they often fail to preserve task-relevant acoustic evidence before reasoning begins. We identify this error pattern as the evidence bottleneck: state-of-the-art systems show larger deficits in acoustic evidence extraction than in downstream reasoning, suggesting that upstream perception is often the limiting factor. To address this problem, we propose EvA (Evidence-First Audio), a dual-path architecture that enhances acoustic evidence preservation through hierarchical aggregation and non-compressive, time-aligned fusion. We also build EvA-Perception, a large-scale training set with about 54K event-ordered captions and 500K evidence-grounded QA pairs. Under a unified zero-shot protocol, EvA achieves the best open-source \emph{Perception} results on MMAU, MMAR, and MMSU, with the largest gains on perception-heavy splits. Human evaluation on open-ended captioning further shows improved fine-grained acoustic coverage and caption quality. These results support the evidence-first hypothesis: stronger audio understanding depends on preserving acoustic evidence before reasoning. Project can be found at \url{https://satsuki2486441738.github.io/EvA/}.
\end{abstract}

\section{Introduction} 
\label{sec:intro}

Large Audio Language Models (LALMs) aim to listen and understand from sound. While recent models like Qwen2.5-Omni~\citep{qwen2.5omni} and Kimi-Audio~\citep{kimiaudio} have demonstrated impressive performance on various benchmarks, their capabilities degrade sharply when confronted with complex acoustic scenes involving fine-grained temporal cues.

\begin{figure}[t]
    \centering
    \includegraphics[width=\columnwidth]{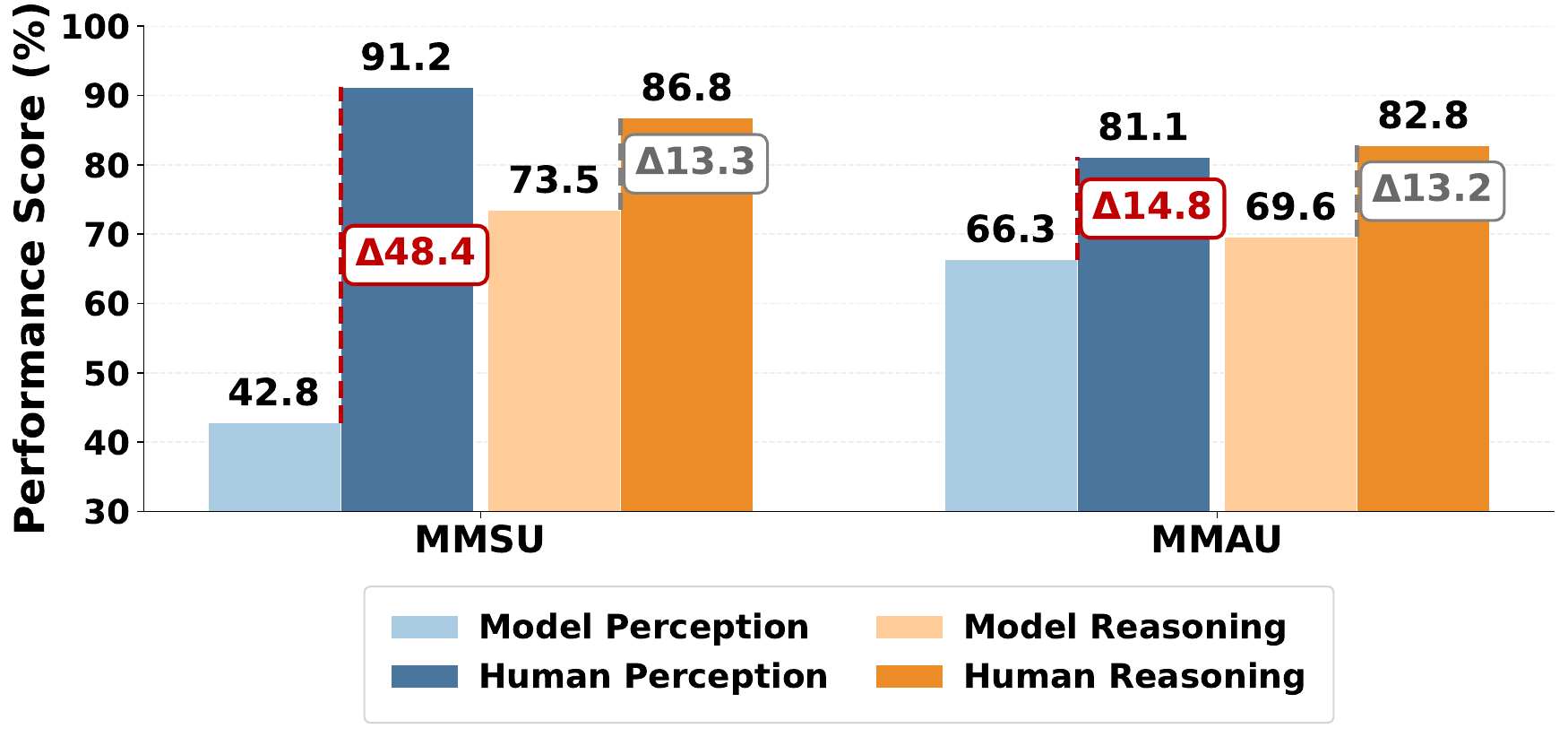}
    \caption{Perception--reasoning performance gap comparison between models and humans. Performance is averaged over Qwen2.5\textendash Omni and Kimi\textendash Audio.}
    \label{fig:perf_gap}
\end{figure}
\noindent We observe a gap between leading LALMs and human performance on perception-oriented tasks, which refer to acoustic entity extraction, and this gap is substantially larger than that on reasoning-oriented tasks, which need inference or composition over the extracted information (Fig.~\ref{fig:perf_gap}). On MMSU, the model--human gap reaches 48.4\% on perception, but only 13.3\% on reasoning. This pattern is counterintuitive: although as a prerequisite for downstream reasoning, perception is often expected to be the easier stage, it appears to be the dominant bottleneck for current LALMs.

\noindent We refer to this pattern as the ``\textbf{evidence bottleneck}'': \textit{current LALMs are often more limited by how well they preserve and represent acoustic information than by how well they reason over that information}. We identify three common design choices that can exacerbate this evidence bottleneck: (1) \textbf{Reasoning-backend overemphasis}: many efforts optimize the downstream reasoning backend through better leverage SFT and RL, while paying less attention to whether the front-end supplies sufficient acoustic evidence. Such optimization can improve how to use acoustic information, but \textbf{cannot recover them} already discarded upstream. (2) \textbf{non-speech information loss}: many LALMs use speech-oriented encoders such as Whisper as the audio front-end. These encoders are optimized to preserve linguistic content, but can compress or discard non-speech acoustic cues that are crucial for general audio understanding. (3) \textbf{Weak alignment interfaces}: existing dual-path systems often rely either on lossy temporal compression, such as Q-Former modules~\citep{salmonn}, or on simple feature concatenation without a shared temporal coordinate~\citep{gama}. Both interfaces can make it difficult for the LLM to jointly use speech and non-speech evidence.

\noindent To address this bottleneck, we introduce \textbf{EvA (Evidence-First Audio)} architecture designed to preserve acoustic evidence before reasoning. EvA uses two complementary streams: a Whisper encoder for speech content and a non-speech encoder (CED-Base~\citep{dinkel2024ced}) for non-speech events. Its core design is a two-stage, non-compressive fusion process. First, EvA builds a dedicated CED evidence stream by aggregating multi-band and multi-layer acoustic representations, so non-speech events are retained across both spectral and representation scales. Second, EvA aligns this non-speech stream to the Whisper timeline and fuses it with the speech stream, combining different cues without temporal compression. These designs keep complementary acoustic evidence accessible to the LLM while preserving the original temporal granularity.

\noindent We further enhance perception learning by developing EvA-Perception, a large-scale audio QA dataset built from temporal annotations in AudioSet-Strong~\citep{audiosetstrong}. It contains 54K temporal and fine-grained captions and 500K QA pairs. Fine-tuned on only 380 hours of audio, EvA achieves strong and consistent gains on complex audio understanding benchmarks, including MMAU~\citep{mmau}, MMAR~\citep{mmar}, MMSU~\citep{mmsu}, and CochlScene~\citep{jeong2022cochlsceneacquisitionacousticscene}. The gains are largest on perception-oriented tasks, which is consistent with our evidence-first hypothesis.

Our main contributions are as follows:
\begin{itemize}
\item \textbf{Diagnosis of ``Evidence Bottleneck''.} We identify an error pattern in current LALMs: the limitation often lies in upstream perception, not downstream reasoning.
\item \textbf{The EvA Architecture.} We propose EvA, a dual-path architecture that preserves acoustic evidence before reasoning through hierarchical aggregation and non-compressive, time-aligned fusion.
\item \textbf{Evidence-Grounded Data and Model.} We release EvA-Perception for training evidence-aware audio perception, together with the EvA model, which achieves competitive Perception results on benchmarks.
\end{itemize}

\section{Related Works}
\label{sec:related_work}
\subsection{Large Audio Language Models}
Large Audio Language Models (LALMs) have progressed rapidly, with models such as Qwen2-Audio~\citep{qwen2-audio}, Qwen2.5-Omni~\citep{qwen2.5omni}, and Kimi-Audio~\citep{kimiaudio}. Most of these models rely on a Whisper encoder~\citep{whisper}, which is effective for speech but less suited to preserving non-speech evidence. Optimized for ASR, Whisper represents spectrograms mainly as a 1D temporal token sequence, which can weaken frequency-localized cues for non-speech events. Prior dual-path designs partly address this issue, but SALMONN~\citep{salmonn} uses lossy Q-Former compression, and GAMA~\citep{gama} combines features without temporal alignment (details in Appendix~\ref{app:qformer}). These limitations motivate our dual-path, non-compressive, time-aligned fusion design.

\subsection{Audio Understanding Benchmarks}
Audio understanding evaluation has moved beyond captioning datasets such as AudioCaps~\citep{audiocaps} and Clotho~\citep{clotho} toward more challenging benchmarks, including MMAU~\citep{mmau}, MMAR~\citep{mmar}, and MMSU~\citep{mmsu}. These benchmarks require both fine-grained perception of acoustic entities as well as higher-level reasoning over such evidence. This makes evidence preservation a prerequisite for reliable audio understanding, motivating our evidence-first perspective and the construction of EvA-Perception.

\begin{figure*}[t]
    \centering
    \includegraphics[width=\textwidth]{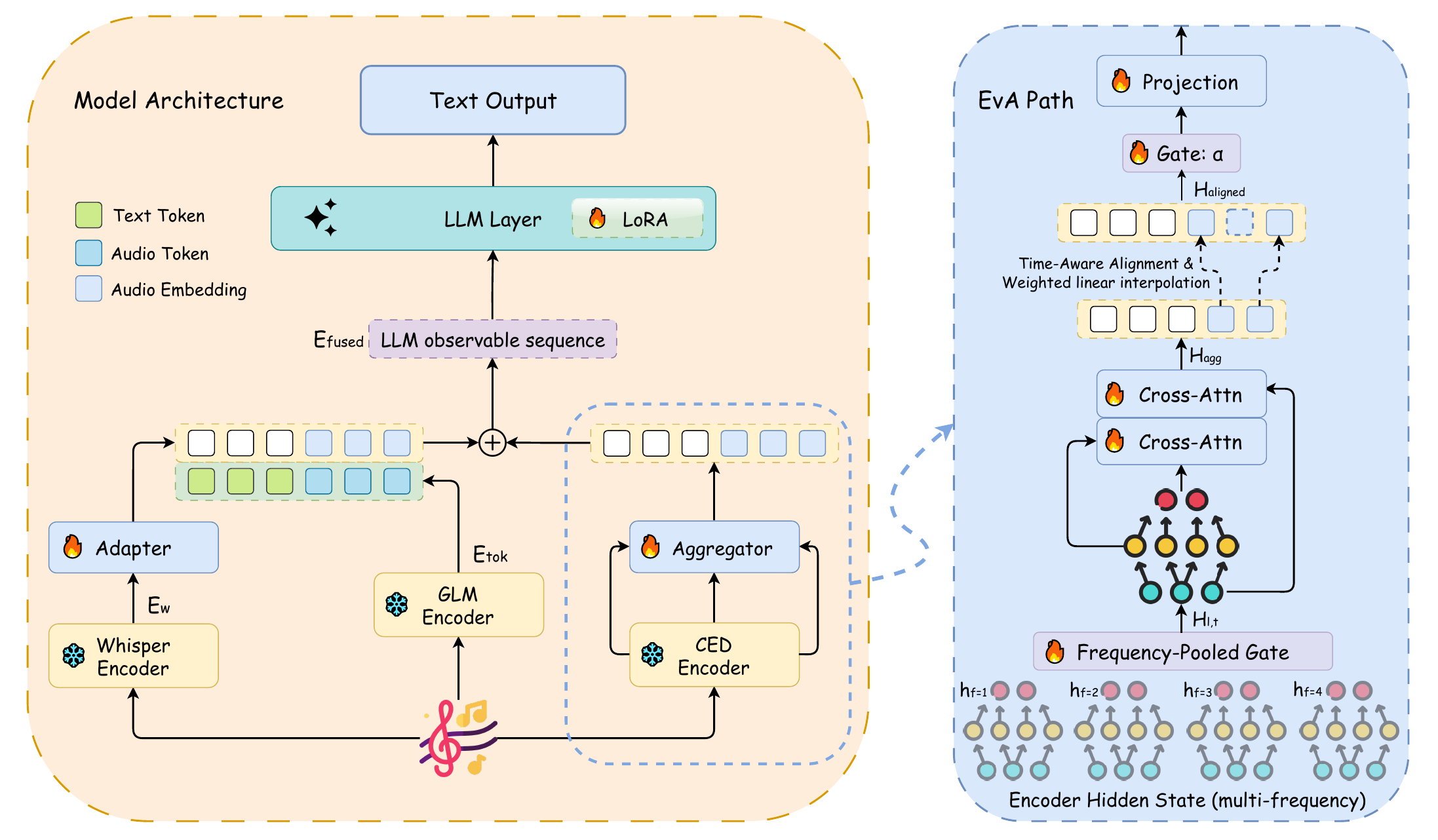}
    \caption{\textbf{Model architecture.} The left half of the left panel shows the Kimi-Audio backbone, while the right half illustrates the additional EvA Path modules. Audio is encoded into four frequency-band features by the CED Encoder, pooled across frequencies, fused via cross-attention, temporally aligned with Whisper, and integrated through gated additive fusion.}
    \label{fig:architecture}
\end{figure*}

\section{Motivation: the Evidence Bottleneck}
We trace the flow of acoustic evidence through LALM inference to identify where it may be attenuated. We first discuss single-path limitations (Sec.~\ref{subsec:single_path_limit}) and then motivate dual-path architectures for preserving complementary evidence (Sec.~\ref{subsec:dual_path_advantage}). Formal derivations are in Appendix~\ref{app:iflow-derivations}.

\paragraph{Notation.}
We use $Z$ to denote the latent acoustic evidence required for the downstream task, such as event identities, temporal boundaries, and event order. Let $X$ be the raw audio waveform, $H$ the encoder hidden representation, $O$ the representation passed to the language model, and $Y$ the model output. We denote mutual information by $I(\cdot;\cdot)$ and assume only that a joint distribution over $(Z, X)$ exists.

\subsection{Single-Path LALMs: An Information Constraint}
\label{subsec:single_path_limit}
Conditioned on fixed parameters after training, the inference-time forward pass is a composition of \emph{deterministic} mappings of $X$:
\[
H = E_{\theta}(X),\qquad O = P_{\theta}(H),\qquad Y = \pi_{\theta}(O).
\]
Under this setting, the data-processing inequality (DPI) for deterministic functions implies
\begin{equation}
I(Z;Y)\;\le\;I(Z;O)\;\le\;I(Z;H)\;\le\;I(Z;X).
\label{eq:dpi}
\end{equation}
We use Eq.~\eqref{eq:dpi} \emph{conceptually} to compare \emph{relative} evidence retention across stages and paths. Post-training that optimizes only the policy $\pi_\theta$ (e.g., SFT or RL) can improve how well the model \emph{uses} evidence already present in $O$, but it does not \emph{restore} acoustic evidence that was lost upstream along $X\!\to\!H\!\to\!O$. This observation motivates architectural choices that preserve acoustic evidence before the LLM (see Sec.~\ref{subsec:implications}). A derivation of \eqref{eq:dpi} under fixed parameters appears in Appendix~\ref{app:iflow-derivations}.

\subsection{Dual-Path Architectures: More Complementary Evidence}
\label{subsec:dual_path_advantage}
Consider two complementary perception paths over the same input $X$: a speech path producing $O_1 = P_1(E_1(X))$ (e.g., Whisper) and a non-speech path producing $O_2 = P_2(E_2(X))$ (e.g., CED-Base). The LLM receives the joint observation $(O_1, O_2)$.

\noindent By the chain rule, we have $I(Z;O_1,O_2) = I(Z;O_1) + I\!\left(Z;O_2\,\middle|\,O_1\right) \ge I(Z;O_1).$ This provides a qualitative motivation for dual-path perception: if the second path captures cues missing from the first, the joint observation can retain complementary acoustic evidence. We examine the empirical effect of adding the CED path through ablations in Table~\ref{tab:ced_ablation}.

\subsection{Design Implications}
\label{subsec:implications}

Taken together, the single-path constraint and dual-path complementarity point to two practical design requirements.

\noindent\textbf{Prioritize the perceptual front-end.}
If acoustic evidence is lost before reaching the LLM, later reasoning modules can only operate on an incomplete representation. Improving the encoder and fusion interface is therefore essential for audio understanding.

\noindent\textbf{Use time-aligned, non-compressive fusion.}
Fusion should preserve temporal resolution and avoid compressing acoustic evidence into a small set of latent tokens. This keeps complementary speech and non-speech cues available without introducing an additional temporal bottleneck.

\section{Method: Evidence-First Audio Understanding Paradigm}
\label{sec:method}
\subsection{Architecture}
EvA adopts a dual-path architecture built on the \textbf{Kimi-Audio-7B} backbone. As shown in Figure~\ref{fig:architecture}, the raw waveform is encoded by two complementary encoders: a Whisper path for speech content and a CED path for non-speech evidence. Their features are aligned to the token timeline and injected into the backbone LLM input space without changing sequence length.

\paragraph{Complementary Dual Encoders.}
We use two encoders that capture complementary acoustic information:
(1) The \textbf{Whisper encoder} ($E_W$) is pre-trained on large-scale ASR corpora and extracts robust linguistic features for speech-related tasks.
(2) The \textbf{CED-Base encoder} ($E_C$) is a Vision Transformer (ViT)-based model trained for general audio event recognition. It captures non-speech cues such as background events and music.
\noindent We extract hidden states from its shallow, middle, and final layers to obtain multi-scale acoustic features. These encoders provide complementary views of the same acoustic scene. In our information flow formulation, they correspond to the two information channels $O_1$ and $O_2$ that feed the downstream LLM.

\paragraph{Hierarchical Evidence Aggregation.}
Standard encoders, which only expose their final-layer features, create an internal information bottleneck long before the LLM. The Data Processing Inequality dictates that these final features cannot be more informative than the collection of intermediate representations. To mitigate this loss, we introduce a hierarchical aggregation process that fuses and harvests features across the frequency domain and from multiple network depths.

\noindent First, in the \textbf{frequency domain}, we leverage the fact that the ViT-based CED encoder's feature maps implicitly retain a frequency axis. For the raw feature maps $\mathbf{\tilde{h}}_l \in \mathbb{R}^{B \times T \times F \times D_c}$ extracted from layer $l \in \{4, 8, L\}$ (where $F$ is the number of frequency bands), we apply a lightweight gated attention mechanism. This performs a learnable, weighted pooling across the frequency bands for each time step:
\begin{equation}
\mathbf{h}_{l,t} = \sum_{f=1}^{F} \mathrm{softmax}(\mathrm{gate}(\mathbf{\tilde{h}}_{l,t,f})) \cdot \mathbf{\tilde{h}}_{l,t,f}
\label{eq:freq_gating}
\end{equation}
This operation dynamically focuses on the most salient frequency bands at each moment, compressing the 2D feature map into a more informative 1D temporal sequence, which we denote as $\mathbf{H}_l \in \mathbb{R}^{B \times T \times D_c}$.

\noindent Second, in the \textbf{cross-layer domain}, we fuse these frequency-aggregated features, $\mathbf{H}_4, \mathbf{H}_8, \text{ and } \mathbf{H}_L$, using a two-stage cascaded cross-attention mechanism implemented in our \textit{Aggregator}. It first enriches the high-level semantic features $\mathbf{H}_L$ with mid-level temporal details from $\mathbf{H}_8$, and then grounds the resulting representation with low-level acoustic patterns from $\mathbf{H}_4$:
\begin{equation}
\mathbf{H}' = \mathrm{Norm}(\mathrm{CrossAttn}(\mathbf{H}_{L}, \mathbf{H}_{8}, \mathbf{H}_{8}) + \mathbf{H}_{L})
\label{eq:cross_attn_1}
\end{equation}
\begin{equation}
\mathbf{H}_{\mathrm{agg}} = \mathrm{Norm}(\mathrm{CrossAttn}(\mathbf{H}', \mathbf{H}_{4}, \mathbf{H}_{4}) + \mathbf{H}')
\label{eq:cross_attn_2}
\end{equation}
Here, $\mathrm{CrossAttn}(\mathbf{Q},\mathbf{K},\mathbf{V})$ denotes cross-attention with query $\mathbf{Q}$, key $\mathbf{K}$, and value $\mathbf{V}$; $\mathrm{Norm}(\cdot)$ denotes layer normalization.
This cascaded, two-stage aggregation process produces a informative feature sequence $\mathbf{H}_{\mathrm{agg}}$ that embodies acoustic evidence integrated across both multiple frequency bands and multiple levels of abstraction.

\begin{table*}[t]
\centering
\resizebox{\textwidth}{!}{  
\begin{tabular}{lccccccc}
\toprule
\textbf{Name} & \textbf{\# of Audio/QA} & \textbf{Avg. Caps Len} & \textbf{Visual} & \textbf{Music} & \textbf{Speech} & \textbf{Integration} & \textbf{Temporal Info}\\
\midrule
AudioCaps~\citep{audiocaps}          & 46k/46k   & 9.03      & \ding{55} & \ding{55} & \ding{55} & \ding{55}  & \ding{55}\\
Clotho~\citep{clotho}          & 5k/5k   & 11.00      & \ding{55} & \ding{55} & \ding{55} & \ding{55}  & \ding{55}\\
LAION-Audio-630K~\citep{laion-audio-630k}          & 630k/630k   &   7.30    & \ding{55} & \ding{55} & \ding{55} & \ding{55}  & \ding{55}\\
WavCaps~\citep{mei2024wavcaps}          & 403k/403k   & 7.80      & \ding{55} & \ding{55} & \ding{55} & \ding{55}  & \ding{55}\\
AudioSetCaps~\citep{bai2025audiosetcaps}  & 1.9M/1.9M  & 28.00        & \ding{55} & \ding{55}  & \ding{55} & \ding{55}  & \ding{55}\\
Auto-ACD~\citep{sun2024auto}  & 1.5M/1.5M  & 18.10        & \ding{51}  & \ding{55}  & \ding{55} & \ding{51}  & \ding{55}\\
CompA-R~\citep{gama}  & 62k/200k  & 18.00     & \ding{51}  & \ding{55}  & \ding{55} & \ding{51}  & \ding{55}\\
FusionAudio-1.2M~\citep{chen2025fusionaudio}  & 1.2M/6M & 47.18 & \ding{51} & \ding{51} & \ding{51} & \ding{51}  & \ding{55}\\

\rowcolor[gray]{0.9}

\textbf{EvA-Caps/QA}  & 54K/500K & \textbf{67.99}  & \ding{51} & \ding{51} & \ding{51} & \ding{51}  & \ding{51} \\
\bottomrule

\end{tabular}
}
\caption{Comparison of open-source audio caption datasets. }
\label{tab:datasets_statistics}
\end{table*}

\paragraph{Time-Aware Alignment and Inject-and-Add Fusion.}
The two encoder paths have different temporal resolutions: Whisper features are aligned to the LLM audio-token timeline with an 80\,ms stride, while the aggregated CED features have an effective stride of 160\,ms. To place non-speech evidence on the same timeline, we upsample $\mathbf{H}_{\mathrm{agg}}$ with time-aware linear interpolation, using mel-frame timestamps to compute weighted averages from neighboring CED positions. This yields $\mathbf{H}_{\text{aligned}} \in \mathbb{R}^{B \times T_w \times D_c}$, a CED evidence sequence aligned with the Whisper tokens. The full interpolation procedure is given in Appendix~\ref{app:interpolation_alg}.

\noindent We then project the Whisper features $\mathbf{E}_W$ and aligned CED features $\mathbf{H}_{\text{aligned}}$ into the LLM hidden space with separate lightweight heads, $\mathrm{Proj}_W$ and $\mathrm{Proj}_C$. Instead of concatenating or compressing the two streams, EvA injects CED evidence locally through gated addition. This keeps the original audio-token sequence length unchanged, requires no modification to the LLM backbone, and lets the model gradually incorporate non-speech evidence through the learnable gate $\alpha$.

\noindent The final fused embedding $\mathbf{E}_{\text{fused}}$ is computed based on a mask $\mathbf{M}$ that identifies audio token positions:
\begin{equation}
\begin{aligned}
\mathbf{E}_{\text{audio}}[i] ={}&
\big(\mathbf{E}_{\text{tok}}[i] + \mathrm{Proj}_W(\mathbf{E}_W[i]) + \alpha \cdot \\
& \mathrm{Proj}_C(\mathbf{H}_{\text{aligned}}[i])\big)\cdot \sqrt{2}
\end{aligned}
\end{equation}
\begin{equation}
\mathbf{E}_{\text{fused}}[i] =
\begin{cases}
\mathbf{E}_{\text{audio}}[i], & \mathbf{M}[i]=1 \\
\mathbf{E}_{\text{tok}}[i], & \mathbf{M}[i]=0
\end{cases}
\label{eq:final_fusion}
\end{equation}
where $\mathbf{E}_{\text{tok}}$ denotes the initial token embedding, $\mathbf{M}$ marks audio-token positions, and $\alpha$ is a learnable scalar gate initialized to a small value. This local additive injection preserves the backbone sequence structure while enriching each audio token with aligned non-speech evidence.

\subsection{EvA-Perception: Evidence-Grounded Dataset}
\label{subsec:eva_perception}
Generic audio captions often lack event order, temporal grounding, and fine-grained acoustic evidence (Table~\ref{tab:datasets_statistics}). We therefore build EvA-Perception, an evidence-grounded audio QA dataset covering sound events, speech, music, temporal relations, and scene-level interactions through event-ordered captions and fine-grained QA pairs.

\paragraph{Construction.}
We first construct two self-built resources: \textbf{EvA-Captions} and \textbf{EvA-QA}. We gather complementary evidence from multiple expert models and cross-check these signals to reduce hallucinated descriptions. AudioSet-Strong~\citep{audiosetstrong} provides time-localized audio labels as acoustic priors. Gemini-2.5-Pro~\citep{comanici2025gemini} converts these annotations into event-ordered captions, while Whisper~\citep{whisper}, OpenMu~\citep{openmu}, and Qwen-2.5-VL-72B~\citep{qwen-vl} supply speech, music, and visual context for disambiguation. QwQ-32B~\citep{qwq32b} then consolidates these signals into coherent audio-focused captions and derives evidence-grounded QA pairs. This process produces \textbf{$\sim$54K} captions over \textbf{150 h} of audio and \textbf{$\sim$500K} QA pairs with a closed/open ratio of $2{:}3$.

\paragraph{Training Data Curation.}
Based on EvA-Captions/EvA-QA and existing high-quality audio-language data, we curate two training mixtures for EvA. \textbf{EvA-Alignment} is used for encoder alignment, focusing on mapping CED evidence into the LLM input space. \textbf{EvA-Perception} is used for instruction fine-tuning, emphasizing fine-grained perception, temporal understanding, and evidence-grounded audio reasoning. Detailed data composition and prompts are provided in Appendix~\ref{app:data_construction}.

\noindent Because one component of the captioning pipeline uses video frames for disambiguation, we conduct a post-hoc visual-leakage audit for EvA-Captions; details are provided in Appendix~\ref{app:visual_leakage}.

\sisetup{
  table-number-alignment = center,
  detect-weight = true
}
\begin{table*}[t]
\centering
\scriptsize
\setlength{\tabcolsep}{3pt}
\resizebox{\textwidth}{!}{%
\begin{tabular}{l c c c c c c c}
\toprule
\multirow{2}{*}{\textbf{Model}} &
\multicolumn{2}{c}{\textbf{MMAU-Clean}} &
\multicolumn{2}{c}{\textbf{MMAR}} &
\multicolumn{2}{c}{\textbf{MMSU}} &
\multirow{2}{*}{\textbf{CochlScene}}\\
&
\multicolumn{1}{c}{Perc.} & \multicolumn{1}{c}{Reas.}
& \multicolumn{1}{c}{Perc.} & \multicolumn{1}{c}{Reas.}
& \multicolumn{1}{c}{Perc.} & \multicolumn{1}{c}{Reas.}
\\
\midrule
Qwen2.5-Omni-7B         & 72.06 & 67.36 & 52.80 & 61.43 & 39.96 & 69.91 & 80.38 \\
Audio-Flamingo-3        & 73.90 & \textbf{71.50} & 59.23 & \textbf{61.64} & 45.63 & 77.86 & 75.57 \\
Step-Audio-2-mini       & 66.54 & 68.58 & 54.58 & 61.01 & 44.36 & \textbf{78.32} & 83.24 \\
Audio-Reasoner          & 65.44 & 63.39 & 44.34 & 38.23 & 40.73 & 68.38 & {--} \\
R1-AQA                  & 72.06 & 62.87 & 48.75 & 50.19 & 41.68 & 71.94 & 76.30 \\
\midrule
Kimi-Audio-7B-Instruct  & 66.18 & 59.59 & 56.79 & 58.72 & 45.47 & 71.85 & 86.17 \\
\textbf{EvA(Kimi-Audio)} 
& \textbf{77.57} {\tiny\textcolor{green!50!black}{+11.39}} 
& 64.17 {\tiny\textcolor{green!50!black}{+4.58}} 
& \textbf{59.79} {\tiny\textcolor{green!50!black}{+3.00}} 
& 59.45 {\tiny\textcolor{green!50!black}{+0.73}} 
& \textbf{47.52} {\tiny\textcolor{green!50!black}{+2.05}} 
& 75.41 {\tiny\textcolor{green!50!black}{+3.56}} 
& \textbf{87.04} {\tiny\textcolor{green!50!black}{+0.87}} \\
\bottomrule
\end{tabular}
}
\setlength{\tabcolsep}{6pt}
\caption{Main results under our unified zero-shot setting. For MMAU, we report the decontaminated split, denoted as MMAU-Clean, to reduce potential source-audio overlap with AudioSet-Strong. Green numbers denote absolute gains over Kimi-Audio-7B-Instruct. Definitions of perception and reasoning are given in Sec.~\ref{sec:setup}; the full MMAU decontamination protocol and original-vs-cleaned results are provided in Appendix~\ref{app:mmau_decontamination}.}
\label{tab:main_results}
\end{table*}

\subsection{Training Strategy}
\label{sec:training} 
We train EvA in two stages: alignment tuning for stable fusion integration, followed by instruction fine-tuning for complex audio understanding.

\paragraph{Backbone Initialization.}
We initialize from the public \textbf{Kimi-Audio-7B} checkpoint and keep the Whisper encoder and CED-Base frozen, so training focuses on the new evidence-fusion modules, lightweight adaptation layers and LLM backbone.

\paragraph{Stage~1: Alignment Tuning.}
In this stage, we train only the newly introduced modules (the CED Aggregator and projection heads) using next-token cross-entropy on text tokens. The goal is to map the CED feature space to the LLM input embedding space without disrupting the model's pre-trained weights.

\paragraph{Stage~2: Instruction Fine-Tuning.}
In this stage, we train the CED Aggregator and Whisper adapter, while updating the LLM backbone through \textbf{LoRA} and keeping both encoders frozen. We continue to use the same text-only cross-entropy objective on the training set.

\section{Experiments}
\label{sec:experiments}
We evaluate EvA on multiple audio understanding benchmarks, reporting both official results and unified \emph{Perception}/\emph{Reasoning} splits. The largest gains appear on perception-oriented tasks, consistent with our evidence-first design.
\subsection{Experimental Setup}
\label{sec:setup}

\paragraph{Benchmarks.}
We evaluate mainly on three audio understanding benchmarks, \textbf{MMAU}, \textbf{MMAR}, and \textbf{MMSU}, as well as an acoustic-scene benchmark, \textbf{CochlScene}. Together, these benchmarks cover perception-oriented audio understanding and reasoning. To directly test our evidence-bottleneck hypothesis, we group benchmark sub-tasks along two axes: \textbf{Perception} and \textbf{Reasoning}. The detailed categorization is provided in Appendix~\ref{app:benchmark}. For completeness, we also report results under each benchmark's original categories and describe our handling of answer ordering in Table~\ref{tab:benchmark_res}.

\paragraph{MMAU-Clean Evaluation.}
To reduce potential source-audio overlap from AudioSet-derived data used by multiple audio-language models, we report MMAU results on a decontaminated split, \textbf{MMAU-Clean}. We construct this split by retrieving each benchmark audio against the AudioSet-Strong training pool with CLAP embeddings and removing likely source duplicates. This removes 149 of 1,000 MMAU samples, including 51 Perception and 98 Reasoning samples. We find manually verified duplicates only in MMAU, while high-similarity pairs in MMAR and MMSU mainly reflect semantic or acoustic similarity rather than identical source audio. Original-vs-cleaned results and the full audit protocol are provided in Appendix~\ref{app:mmau_decontamination}.

\paragraph{Baselines.} We compare to strong general understanding and reasoning systems: \textit{Qwen2-Audio}~\citep{qwen2-audio}, \textit{Qwen2.5-Omni}~\citep{qwen2.5omni}, Audio-Flamingo-3~\citep{goel2025audioflamingo3advancing}, Step-Audio-2-mini~\citep{wu2025stepaudio2technicalreport}, \textit{Kimi-Audio}~\citep{kimiaudio}, \textit{Audio-Reasoner}~\citep{xie2025audio}, and \textit{R1-AQA}~\citep{li2025reinforcement}.
All baselines are run under a unified inference protocol, and numbers reported are from our reproduction to ensure fairness.

\paragraph{Implementation Details.}
\label{para:setting}
We follow the two-stage training strategy described in Sec.~\ref{sec:training}. In Stage~1, we train on EvA-Alignment and update only the CED Aggregator while keeping the pre-trained encoders and LLM frozen. In Stage~2, we use EvA-Perception for instruction tuning, update the CED Aggregator and Whisper adapter, and fine-tune the LLM with LoRA. During evaluation, we use greedy decoding with a maximum generation length of $1024$. Detailed hyperparameters, batch sizes, schedules, and hardware settings are provided in Appendix~\ref{app:hyperparameters}.

\begin{figure*}[t]
    \centering
    \includegraphics[width=\textwidth]{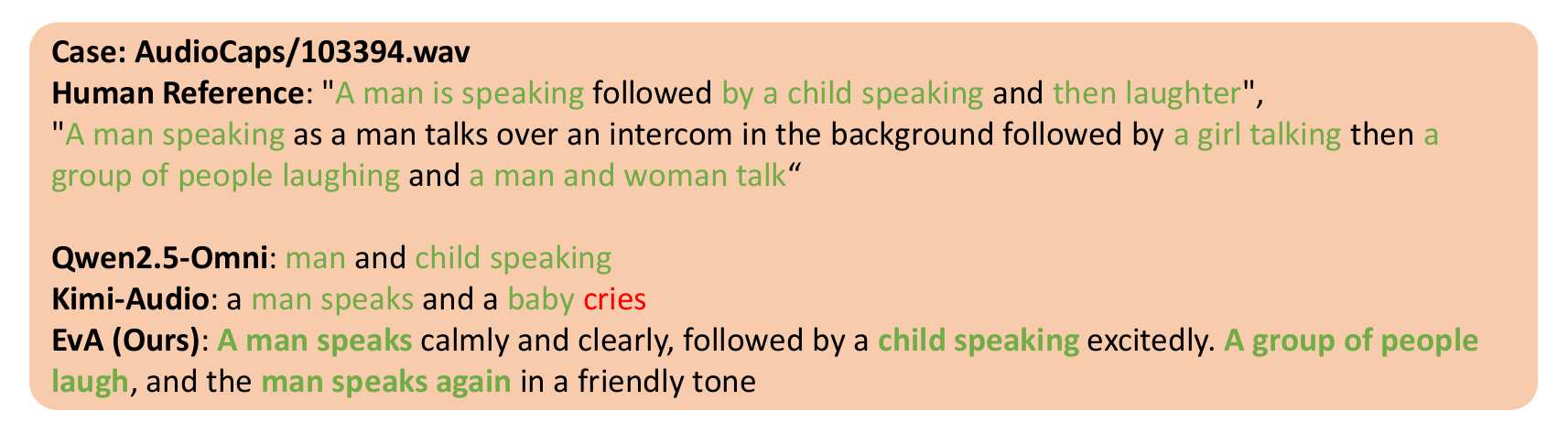}
    \caption{Qualitative comparison of captions generated by different models on AudioCaps examples.}
    \label{fig:case_study}
\end{figure*}

\begin{table*}[t]
\centering
\footnotesize
\resizebox{\textwidth}{!}{%
\begin{tabular}{l lll *{3}{S[table-format=2.2]}}
\toprule
\multirow{2}{*}{\textbf{Stage}} &
\multirow{2}{*}{\textbf{Setting}} &
\multirow{2}{*}{\textbf{Trainables}} &
\multirow{2}{*}{\textbf{Start Ckpt}} &
\multicolumn{3}{c}{\textbf{AudioCaps (CLAP)}} \\
 & & & & \multicolumn{1}{c}{Cos} & \multicolumn{1}{c}{R@1} & \multicolumn{1}{c}{R@5} \\
\midrule
S0     & Base Model            & N/A                    & N/A & 14.61 & 9.27  & 24.97 \\ 
S1(1)  & w/o CED path          & Adapter                & S0  & 35.40 & 18.50 & 43.60 \\
S1(2)  & w/o frequency pooling & Adapter \& CED Agg.    & S0  & 35.54 & 21.24 & 49.61 \\
S1(3)  & w/o crossing fusion   & Adapter \& CED Agg.    & S0  & 28.63 & 11.82 & 29.74 \\
S1(4)  & Q-former              & Adapter \& CED Q-former & S0 & 36.24 & 20.08 & 47.36 \\
S1(0)  & \textbf{EvA Dual-Path} & Adapter \& CED Agg.    & S0  & \bfseries 36.77 & \bfseries 22.77 & \bfseries 49.81 \\
    
\midrule
\midrule
\textbf{Stage} &
\textbf{Setting} &
\textbf{Trainables} &
\textbf{Start Ckpt} &
\textbf{MMAU} &
\textbf{MMAR} &
\textbf{MMSU} \\
\midrule
S0    & Base Model            & N/A                         & N/A   & 65.33 & 49.21 & 43.36 \\
S2(1) & w/o CED path           & Adapter \& LLM               & S1(1) & 66.10 & 47.50 & 54.60 \\
S2(0) & \textbf{EvA Dual-Path} & Adapter \& CED Agg. \& LLM   & S1(0) & \bfseries 67.70 & \bfseries 51.50 & \bfseries 58.10 \\
\bottomrule
\end{tabular}
}
\caption{Ablations of the EvA fusion path. On \textit{AudioCaps}, we report CLAP-based cosine similarity (\textbf{Cos}) and retrieval; on MMAU/MMAR/MMSU, we report \emph{Perception} accuracy. \textit{Adapter} denotes the Whisper adapter, and \textit{LLM} denotes LoRA on the backbone. In Stage~2, \textit{w/o CED path} is trained and evaluated without the CED branch.}
\label{tab:ced_ablation}
\end{table*}

\subsection{Main Results}
\label{sec:main_results}
\paragraph{Results on Understanding Tasks.} As shown in Table~\ref{tab:main_results}, EvA achieves \textbf{77.57/64.17} on MMAU-Clean, \textbf{59.79/59.45} on MMAR, \textbf{47.52/75.41} on MMSU, and \textbf{87.04} on CochlScene. Compared with the Kimi-Audio-7B-Instruct backbone, EvA improves every reported metric, with gains of +11.39/+4.58 on MMAU-Clean, +3.00/+0.73 on MMAR, +2.05/+3.56 on MMSU, and +0.87 on CochlScene. The gains are most pronounced on perception-oriented splits, consistent with our evidence-first design. Notably, EvA remains the best-performing model on the decontaminated MMAU-Clean Perception split, suggesting that its advantage is not simply due to source-audio overlap. Its competitive CochlScene result further indicates that the same design transfers to specialized acoustic-scene recognition.

\paragraph{Case Study}
We illustrate these differences with AudioCaps case studies (Fig.~\ref{fig:case_study}). Compared with Qwen2.5-Omni and Kimi-Audio, \textbf{EvA} produces captions that are more faithful to the audio events. In the example shown, EvA captures the sequence of tone shifts (calm speech $\to$ child excitement $\to$ laughter), whereas the baselines misinterpret the events. This example illustrates how stronger acoustic evidence can improve caption accuracy.
\label{app:qualitative_analysis}

\subsection{Human Evaluation}
\label{sec:human_eval}

We conduct a human evaluation on AudioSet-Strong test-set captioning, where each model is asked to generate an open-ended audio description. Human annotators evaluate three complementary aspects: (i) \textbf{Label Recall}, whether the caption semantically covers the official AudioSet labels; (ii) \textbf{Acoustic Object Perception}, whether the model captures fine-grained audible objects beyond the coarse official labels; and (iii) \textbf{Blind Pairwise Arena}, where annotators compare two anonymous captions for the same audio. These experiments evaluate official-label coverage, fine-grained acoustic perception beyond coarse annotations, and overall audio understanding reflected in open-ended captioning.

\begin{table}[h]
\centering
\small
\setlength{\tabcolsep}{3.5pt}
\resizebox{\columnwidth}{!}{
\begin{tabular}{lcccc}
\toprule
\textbf{Model} & \textbf{Label Recall} & \textbf{\# Obj. Mentions} & \textbf{Obj. Acc.} & \textbf{Arena} \\
\midrule
Step-Audio-2-mini      & 58.6 & 191 & 67.5 & 43.2 \\
Qwen2.5-Omni           & 57.9 & 166 & 78.3 & 55.1 \\
Audio-Reasoner         & 51.9 & 128 & \textbf{85.9} & 64.2 \\
Audio-Flamingo-3       & 50.4 & 125 & 83.2 & 39.7 \\
R1-AQA                 & 47.4 & 123 & 84.6 & 39.1 \\
Kimi-Audio-7B          & 31.6 & 80  & 71.3 & 25.3 \\
\textbf{EvA}           & \textbf{60.2} {\tiny\textcolor{green!50!black}{+28.6}} & \textbf{208} {\tiny\textcolor{green!50!black}{+128}} & 84.1 {\tiny\textcolor{green!50!black}{+12.8}} & \textbf{80.5} \\
\bottomrule
\end{tabular}
}
\caption{Human evaluation summary on AudioSet-Strong captioning. Green numbers denote gains over Kimi-Audio-7B. Full results are in Appendix~\ref{app:human_eval_details}.}
\label{tab:human_eval_summary}
\end{table}

\noindent As shown in Table~\ref{tab:human_eval_summary}, we evaluate captioning quality along three axes: fine-grained coverage, perceptual accuracy, and overall captioning quality. EvA achieves the best label recall and produces the most acoustic-object mentions, indicating stronger fine-grained coverage. It also maintains high object accuracy and obtains the highest blind arena score, suggesting that the additional acoustic evidence improves caption quality rather than merely increasing verbosity.

\subsection{Ablation Studies}

\paragraph{Setup.}
We evaluate the CED fusion path at two levels: its overall contribution during alignment and perception SFT, and the effect of its frequency-gated pooling and top--down cross-layer fusion. Stage~1 variants use \textbf{EvA-Alignment}; Stage~2 compares single- and dual-path variants trained on the same \textbf{EvA-Perception} subset.

\paragraph{Overall Effect of the CED Branch.}
On \textit{AudioCaps}, EvA Dual-Path improves over the \textit{w/o CED path} baseline in CLAP Cos/R@1/R@5, from 35.40/18.50/43.60 to 36.77/22.77/49.81.
This indicates that the CED branch already provides complementary acoustic evidence during alignment.
Under the same-data Stage~2 comparison, EvA Dual-Path further improves perception accuracy over the single-path baseline by +1.60/+4.00/+3.50 on MMAU/MMAR/MMSU.
These gains support the effectiveness of introducing the CED branch throughout training and inference.

\paragraph{Ablation of Modules in the CED Branch.}
Removing top--down cross-layer fusion (S1(3)) causes a clear drop across all CLAP metrics, showing that intermediate CED layers contain useful acoustic evidence beyond the final layer.
Replacing the hierarchical Aggregator with a window-level Q-Former (S1(4)) also underperforms full EvA, supporting our non-compressive multi-level fusion design; Appendix~\ref{app:qformer} provides a structural comparison.
Finally, removing frequency-gated pooling (S1(2)) slightly improves R@5 but lowers Cos and R@1, suggesting that the gate mainly benefits top-ranked alignment quality.
Overall, the ablations support both the CED branch itself and the proposed hierarchical Aggregator.

\subsection{Validation on Other Architecture}
\label{validation}

\begin{table}[h]
\centering
\small
\setlength{\tabcolsep}{5pt}
\begin{tabular}{llcc}
\toprule
Benchmark & Subset & Base Arch. & EvA \\
\midrule
\multirow{2}{*}{MMAU-mini}
& Perception & 75.23 & \textbf{76.16} {\tiny\textcolor{green!50!black}{+0.93}} \\
& Reasoning  & 67.80 & \textbf{69.73} {\tiny\textcolor{green!50!black}{+1.93}} \\
\midrule
\multirow{2}{*}{MMAR}
& Perception & 53.24 & \textbf{56.82} {\tiny\textcolor{green!50!black}{+3.58}} \\
& Reasoning  & 61.61 & \textbf{65.08} {\tiny\textcolor{green!50!black}{+3.47}} \\
\midrule
\multirow{2}{*}{MMSU}
& Perception & 42.02 & \textbf{49.21} {\tiny\textcolor{green!50!black}{+7.19}} \\
& Reasoning  & 69.95 & \textbf{79.47} {\tiny\textcolor{green!50!black}{+9.52}} \\
\bottomrule
\end{tabular}
\caption{The performance of Qwen2.5-Omni architecture.}
\label{tab:qwen25omni-eva-pr}
\end{table}

To further validate the generality of EvA, we apply the same design to Qwen2.5-Omni-7B, using 10\% of EvA-Perception for convenience. As shown in Table~\ref{tab:qwen25omni-eva-pr}, EvA consistently improves performance across all three benchmarks and both perception- and reasoning-oriented subsets. These results indicate that the EvA paradigm has the potential to be transferred to a different LALM and still provide consistent benefits, supporting the feasibility and architectural generality of our approach.

\section{Conclusion}
\label{sec:conclusion}

In this work, we identified the evidence bottleneck as a critical limitation in Large Audio Language Models (LALMs): performance in complex acoustic scenes is often limited more by upstream perception than by downstream reasoning. To address this limitation, we introduced EvA, a dual-path architecture that preserves acoustic evidence through hierarchical aggregation and non-compressive, time-aligned fusion. Supported by the EvA-Perception dataset, EvA achieves strong performance on benchmarks such as MMAU, MMAR, and MMSU, with the largest gains concentrated on perception-heavy tasks. These results are consistent with our central claim that stronger audio understanding depends on preserving acoustic evidence before reasoning.

\section{Limitations}
\label{sec:limitations}
While EvA advances audio understanding with stronger acoustic evidence, several limitations remain: our curated audio--text training corpus currently uses English-only captions, even though the paired audio and evaluation benchmarks contain multilingual inputs, so more systematic multilingual supervision and evaluation are still needed; temporal reasoning is constrained by the soft event boundaries in AudioSet-Strong, and music analysis lacks expert-level concepts such as pitch or harmony. Addressing these challenges is an important direction for future work.

\bibliography{custom}

\clearpage
\appendix

\section{Appendix}

\subsection{Potential Risk}
Because the model and training resources are built from publicly available online content, EvA may inherit latent moral, ethical, or racial biases. We therefore recommend strengthened human review and supervision when using the system, especially in high-stakes or sensitive scenarios.

\subsection{Access Statement}
All training and evaluation data used in this work are collected from publicly available research datasets and open-source resources. We do not use private user recordings or non-public personal data. We cite the creators of external artifacts used in this work, including datasets, benchmarks, pretrained models, audio encoders, and evaluation tools. These artifacts are used for research, evaluation, and model-development purposes, following their intended use when specified by the original creators. EvA-Perception and released derivatives are intended for research and reproducibility contexts, and should respect the access conditions of the original artifacts.

\subsection{Artifact Documentation}
We document EvA-Perception and EvA-Alignment in Sec.~\ref{subsec:eva_perception} and Appendix~\ref{app:data_construction}. The artifacts cover sound events, speech, music, temporal relations, scene-level interactions, event-ordered captions, and evidence-grounded QA. Table~\ref{tab:dataset_composition} reports constituent sources, modalities, and quantities. The released captions and QA pairs are in English; limitations in multilingual coverage, temporal boundary precision, and expert-level music concepts are discussed in Sec.~\ref{sec:limitations}.

\subsection{Use of LLMs}
Large language models were utilized for specific tasks, such as assisting with coding and providing grammar checks and language refinement during the writing of this paper. All scientific content, including research design, experimentation, data analysis, and conclusions, was independently conducted by the authors without LLM involvement in the core research process.

\subsection{Human Annotation Statement}
We recruited five unpaid volunteer annotators with at least undergraduate-level education for a small-scale human evaluation. Annotators were asked to judge whether model captions recall AudioSet labels, whether mentioned acoustic objects are audible in the input, and which caption is better in blind pairwise comparisons, with ties allowed. The evaluation used public benchmark audio and model-generated captions only, involved no private personal data, and posed minimal risk to participants. Volunteers were informed of the annotation tasks and could stop participation at any time.

\subsection{Notes for the Information-Flow View}
\label{app:iflow-derivations}

\paragraph{DPI under deterministic inference.}
After training, condition on fixed parameters $\theta$.
Let $H=E_\theta(X)$, $O=P_\theta(H)$, $Y=\pi_\theta(O)$ be deterministic mappings of $X$.
For any measurable $f$, DPI gives $I(Z;f(X))\le I(Z;X)$.
Applied stage-wise to the composition,
\[
I(Z;Y)\le I(Z;O)\le I(Z;H)\le I(Z;X).
\]
If inference includes independent randomness $U$ (e.g., stochastic decoding), write $Y=g(O,U)$ with $U\!\perp\! Z\,|\,O$, so $I(Z;Y)\le I(Z;O,U)=I(Z;O)$.

\paragraph{Chain-rule identity (for intuition).}
For random variables $Z,O_1,O_2$, the chain rule gives
$
I(Z;O_1,O_2)=I(Z;O_1)+I\!\left(Z;O_2\,\middle|\,O_1\right)\ge I(Z;O_1).
$
We use this only to express complementarity succinctly; no quantitative claim is made.

\subsection{Algorithm for Time-Aware Coverage-Weighted Linear Interpolation}
\label{app:interpolation_alg}

\begin{algorithm*}[t]
\caption{Time-Aware, Coverage-Weighted Linear Interpolation (0-based indexing)}
\label{alg:interpolation}
\begin{algorithmic}[1]
\Require Aggregated CED features $\mathbf{H}_{\mathrm{agg}} \in \mathbb{R}^{T_c \times D}$
\Require CED centers $t_c[0..T_c-1]$ (monotonic), Whisper centers $t_w[0..T_w-1]$ (monotonic)
\Require Coverage weights $c[0..T_c-1]$, with $c[\ell]\in[0,1]$
\Require Stability constant $\varepsilon > 0$ (e.g., $10^{-8}$)
\Ensure Aligned features $\mathbf{H}_{\mathrm{aligned}} \in \mathbb{R}^{T_w \times D}$
\State Initialize $\mathbf{H}_{\mathrm{aligned}}$ as a tensor of shape $(T_w, D)$
\If{$T_c \le 1$} \State \Return $\mathbf{H}_{\mathrm{aligned}} \leftarrow$ repeat the sole vector to length $T_w$ \EndIf
\For{$k \leftarrow 0$ to $T_w-1$}
    \State \Comment{Locate neighbors of $t_w[k]$ in $t_c$ via binary search}
    \State $r \leftarrow \mathrm{searchsorted}(t_c,\, t_w[k])$
    \State $r \leftarrow \mathrm{clamp}(r,\, 1,\, T_c-1)$,\quad $l \leftarrow r - 1$
    \State \Comment{Linear interpolation factor with clamping}
    \State $\alpha \leftarrow \dfrac{t_w[k] - t_c[l]}{t_c[r] - t_c[l] + \varepsilon}$;
    \State \quad $\alpha \leftarrow \mathrm{clamp}(\alpha,\,0,\,1)$
    \State \Comment{Coverage-weighted, normalized interpolation}
    \State $\mathbf{x}_L, \mathbf{x}_R \leftarrow \mathbf{H}_{\mathrm{agg}}[l],\ \mathbf{H}_{\mathrm{agg}}[r]$
    \State $c_L, c_R \leftarrow c[l],\ c[r]$
    \State $\text{num} \leftarrow (1-\alpha)\,(c_L\,\mathbf{x}_L)\ +\ \alpha\,(c_R\,\mathbf{x}_R)$
    \State $\text{den} \leftarrow (1-\alpha)\,c_L\ +\ \alpha\,c_R\ +\ \varepsilon$
    \State $\mathbf{H}_{\mathrm{aligned}}[k] \leftarrow \text{num} / \text{den}$
\EndFor
\State \Return $\mathbf{H}_{\mathrm{aligned}}$
\end{algorithmic}
\end{algorithm*}

\paragraph{Assumptions and Notation.}
We denote the aggregated CED sequence as $\mathbf{H}_{\mathrm{agg}} \in \mathbb{R}^{T_c \times D}$ with per-feature centers $t_c[0],\dots,t_c[T_c-1]$ (monotonically increasing), and the target Whisper centers $t_w[0],\dots,t_w[T_w-1]$ (also monotonic). All timestamps share the same unit (e.g., mel frames or milliseconds). A small constant $\varepsilon>0$ (we use $10^{-8}$) is added for numerical stability. When $T_c \le 1$, we simply repeat the sole vector to length $T_w$.

\paragraph{Coverage weights.}
For each CED window $\ell$, its coverage weight $c[\ell]\in[0,1]$ measures the fraction of the window overlapping valid (non-padded) audio. Concretely, if a window starts at $\text{start}_\ell$ and ends at $\text{end}_\ell$ with window size $t_{\mathrm{sz}}$, and the valid audio spans $[0, T_{\mathrm{mel}}-1]$, then
\[
c[\ell] \;=\; \frac{\max\!\big(0,\,\min(\text{end}_\ell,\,T_{\mathrm{mel}}-1) - \text{start}_\ell + 1\big)}{t_{\mathrm{sz}}}\,.
\]
Thus $c=1$ for fully valid windows and decreases as padding overlap grows.

\paragraph{Target centers for Whisper.}
Let $step_{mel}$ be the mel frames per Whisper token and $center_{mel}$ its center offset. For $k=0,\dots,T_w-1$, we use
\begin{equation}
    t_w[k] = k \cdot step_{mel} + center_{mel}.
    \label{eq:whisper-centers}
\end{equation}
In our implementation, $step_{mel}=8$ and $center_{mel}=4$.

\paragraph{Discussion.}
By reweighting both neighbors with $c[\ell]$ in the numerator \emph{and} renormalizing by the weighted sum in the denominator, features largely sourced from padded/silent regions contribute less to the aligned representation, especially near boundaries. In practice we use a vectorized implementation (single search-sorted, broad-casted arithmetic) that avoids Python loops while preserving the above semantics.

\subsection{Training setting}
As shown in Table~\ref{tab:hyperparams}, we summarize the training hyperparameters for both stages. Notably, the CED Aggregator is trained from scratch in Stage~1, while the Whisper adapter and LLM are only updated in Stage~2. Both stages use a cosine learning rate schedule with a small warmup, and we keep the encoders frozen throughout to preserve their pre-trained representations and ensure stable training of the new modules. The LoRA configuration for the LLM is chosen to balance expressiveness with parameter efficiency, targeting the query, key, value, and output projection matrices without modifying the MLP layers. We also prepare for DeepSpeed ZeRO-3 optimization but keep it off by default to allow flexibility in resource-constrained settings.
\label{app:hyperparameters}
\begin{table*}[!htbp]
\centering
\resizebox{\textwidth}{!}{
\begin{tabular}{lcc}
\toprule
 & \textbf{Stage~1 (Alignment)} & \textbf{Stage~2 (SFT, LoRA)} \\
\midrule
Trainable modules & CED Aggregator & CED Aggregator, Whisper adapter; LLM via LoRA \\
Dataset & EvA-Alignment & EvA-Perception \\
Epochs & 5 & 2 \\
Per-device batch & 2 & 2 \\
Grad. accumulation & 8 & 16 \\
Global batch (8$\times$A100) & 128 & 256 \\
Optimizer & AdamW ($\beta_2{=}0.95$, wd $0.1$) & AdamW ($\beta_2{=}0.95$, wd $0.1$) \\
LR / schedule / warmup & $1\!\times\!10^{-3}$ / cosine / 1\% & $5\!\times\!10^{-5}$ / cosine / 1\% \\
Max seq length & 512 & 1024 \\
LoRA (LLM) & -- & $r{=}64,\ \alpha{=}64,\ \text{dropout}=0.05$; targets=\texttt{q,k,v,o} (\texttt{include\_mlp}=False) \\
Modules to save & -- & \texttt{model.vq\_adaptor} \\
Checkpoint export & split every epoch (keep last 5) & split every epoch (keep last 5) \\
Distributed setup & torchrun, 8$\times$A100-80GB & torchrun, 8$\times$A100-80GB \\
DeepSpeed ZeRO-3 & prepared (config available), \emph{off} by default & prepared, \emph{off} by default \\
Runtime (wall-clock) & $\sim$12h & $\sim$12h \\
\bottomrule
\end{tabular}}
\caption{Training hyperparameters. Encoders (Whisper, CED-Base) are frozen throughout.}
\label{tab:hyperparams}
\end{table*}

\subsection{Data Construction}
\label{app:data_construction}
Figure~\ref{fig:caption_generation} and Figure~\ref{fig:qa_generation} show the instructions used to construct EvA-Perception. These instructions are designed to produce audio-focused descriptions while controlling cross-modal interference and marking ambiguities based only on auditory perception. The input sources include audio tags, audio descriptions, speech content (ASR), music descriptions, and video descriptions, each with specific guidelines on how to use them for accurate audio captioning. The processing steps outline a systematic approach to multimodal parsing, auditory fact determination, ambiguity inference, reliability assessment, and final caption generation.

\subsection{Cases in EvA-Captions and EvA-QA}
We provide two cases from our EvA-Captions and EvA-QA datasets in Figure~\ref{fig:case1} and Figure~\ref{fig:case2}, respectively. These cases illustrate the complexity and richness of the audio descriptions and the corresponding QA pairs, showcasing the model's ability to handle intricate acoustic scenes and extract meaningful information for question answering tasks. The captions contain detailed descriptions of the audio content, while the QA pairs test various aspects of understanding, such as emotional effects, vocal presence, sound dominance, and the contribution of specific elements to the overall atmosphere.

\subsection{Visual-Leakage Audit}
\label{app:visual_leakage}

Since Qwen2.5-VL-72B is used in our caption-generation pipeline, generated captions may occasionally contain information derived from video frames rather than audio alone. We therefore conduct a post-hoc audit to estimate potential visual leakage in EvA-Captions.

\paragraph{Setup.}
We randomly sample 500 captions from EvA-Captions using seed 42. Each caption is judged independently by an LLM-based auditor. The auditor is given only the generated caption and is asked to determine whether it contains information that would require visual evidence rather than audio evidence.

\paragraph{Rubric.}
A caption is flagged if it contains visible-only details such as color, physical appearance, clothing, lighting, spatial layout, on-screen text, brand logos, or other visual scene information. Descriptions of audible sources, such as ``dog barking'', ``car engine'', ``piano'', or ``applause'', are not considered leakage if they are plausibly inferable from audio.

\paragraph{Results.}
The audit flags 7 out of 500 captions as potentially containing visual-only information, corresponding to a conservative leakage estimate of 1.4\%. Representative flagged phrases include ``vintage-style telephone bell'', ``indoor environment'', ``small dog'', and ``outdoor nighttime ambiance''. Some of these phrases may be partially inferable from audio, but we count them as potential leakage under the conservative rubric.

\begin{table}[t]
\centering
\small
\begin{tabular}{ll}
\toprule
\textbf{Item} & \textbf{Result} \\
\midrule
Sample size & 500 captions \\
Sampling seed & 42 \\
Audit input & Generated caption only \\
Flagged captions & 7 / 500 \\
Estimated leakage rate & 1.4\% \\
\bottomrule
\end{tabular}
\caption{Summary of the visual-leakage audit.}
\label{tab:visual_leakage_summary}
\end{table}

\paragraph{Limitation.}
This audit suggests that visual leakage exists but occurs at a low rate in the sampled captions. It does not prove that EvA-Captions is fully free of visual-only information, nor does it remove such cases from the training data. Future versions of the pipeline can incorporate an automatic filtering or rewriting stage to further reduce visual-only descriptions.

\subsection{Human Evaluation Details}
\label{app:human_eval_details}

Table~\ref{tab:human_eval_details} provides the raw counts behind the summary in Table~\ref{tab:human_eval_summary}. The evaluation covers three aspects: fine-grained coverage, measured by Label Recall and \# Obj. Mentions; perceptual accuracy, measured by Obj. Acc.; and overall captioning quality, measured by Blind Pairwise Arena. In the arena setting, annotators compare anonymous captions for the same audio, with ties counted as half a win.

\begin{table*}[t]
\centering
\small
\setlength{\tabcolsep}{4pt}
\resizebox{\textwidth}{!}{
\begin{tabular}{lcccccc}
\toprule
\textbf{Model} & \textbf{Label Recall} & \textbf{\# Obj. Mentions} & \textbf{Obj. Acc.} & \textbf{Arena Shown} & \textbf{W/L/T} & \textbf{Arena} \\
\midrule
Step-Audio-2-mini      & 78 / 133 (58.6) & 191 & 129 / 191 (67.5) & 154 & 50 / 71 / 33  & 43.2 \\
Qwen2.5-Omni           & 77 / 133 (57.9) & 166 & 130 / 166 (78.3) & 138 & 63 / 49 / 26  & 55.1 \\
Audio-Reasoner         & 69 / 133 (51.9) & 128 & 110 / 128 (85.9) & 169 & 94 / 46 / 29  & 64.2 \\
Audio-Flamingo-3       & 67 / 133 (50.4) & 125 & 104 / 125 (83.2) & 165 & 46 / 80 / 39  & 39.7 \\
R1-AQA                 & 63 / 133 (47.4) & 123 & 104 / 123 (84.6) & 137 & 36 / 66 / 35  & 39.1 \\
Kimi-Audio-7B          & 42 / 133 (31.6) & 80  & 57 / 80 (71.3)   & 162 & 23 / 103 / 36 & 25.3 \\
\textbf{EvA}           & \textbf{80 / 133 (60.2)} & \textbf{208} & 175 / 208 (84.1) & 169 & 122 / 19 / 28 & \textbf{80.5} \\
\bottomrule
\end{tabular}
}
\caption{Detailed human evaluation on AudioSet-Strong captioning. Values in parentheses are percentages; Arena treats ties as half a win.}
\label{tab:human_eval_details}
\end{table*}

\subsection{Introduction on Benchmarks}
\label{app:benchmark}

\textbf{MMAU} covers broad audio modalities with a mixture of perceptual, information-extraction, and reasoning questions. 
\textbf{MMAR} emphasizes multi-step inference across hierarchical layers. 
\textbf{MMSU} targets spoken language understanding, including fine-grained linguistic and paralinguistic phenomena. 

For our unified analysis, we keep \textit{MMSU}'s native tags, map \textit{MMAU}'s \emph{information extraction}/\emph{reasoning} categories to Perception/Reasoning, and map \textit{MMAR}'s \emph{Signal+Perception}/\emph{Semantic+Culture} categories in the same way. 

In the original MMAU test set, the reference answers are distributed unevenly across choice positions: 
\textbf{A: 39.5\%, B: 27.1\%, C: 20.8\%, and D: 12.6\%. }
Such imbalance may bias evaluation for models with positional preferences (e.g., favoring earlier options). 
To mitigate this artifact, we randomized the order of choices, ensuring that the final distribution of correct answers is balanced across positions. 
All reported MMAU results in the main paper are based on this balanced setting.

\subsection{MMAU Decontamination Audit}
\label{app:mmau_decontamination}

\paragraph{Motivation.}
MMAU contains AudioSet-derived audio, and AudioSet is widely used in audio-language model pretraining and instruction tuning. Since the training data of several evaluated systems are not fully public, possible source-audio overlap may affect multiple models rather than only EvA. We therefore construct a decontaminated MMAU split, denoted as MMAU-Clean, and use it as the main MMAU evaluation setting in Table~\ref{tab:main_results}.

\paragraph{Audit scope.}
Our goal is to detect likely source-audio overlap, rather than broad semantic or acoustic similarity. In other words, we do not remove benchmark samples merely because they contain similar sound events. We only remove samples that are likely to share the same or near-duplicate source audio with the AudioSet-Strong clips used in EvA-Perception.

\paragraph{Protocol.}
We build an embedding registry for all 79,555 AudioSet-Strong training clips used in EvA-Perception. For each sample in MMAU, MMAR, and MMSU, we compute an audio embedding using LAION CLAP \texttt{laion/clap-htsat-unfused}, and retrieve its nearest AudioSet-Strong training clip by cosine similarity.

The removal threshold is determined by manual verification. We sort benchmark--training pairs by CLAP similarity, divide them into similarity buckets, and manually inspect audio pairs in each bucket. We find that pairs below 0.80 CLAP cosine similarity do not show consistent source-audio duplication, while genuine duplicate audio in MMAU appears in the $\geq 0.80$ range. We therefore use top-1 CLAP cosine similarity $\geq 0.80$ as a conservative removal criterion. This threshold is recall-oriented: it may remove some acoustically similar but non-identical samples, but it reduces the chance of retaining duplicated source audio.

\paragraph{Audit result.}
Applying this protocol to MMAU, MMAR, and MMSU, we find manually verified source-audio duplicates only in MMAU. For MMAR and MMSU, high-similarity pairs are caused by acoustic or semantic similarity rather than identical source audio, so we do not remove samples from these two benchmarks.

For MMAU, we remove 149 out of 1,000 samples using the CLAP similarity $\geq 0.80$ criterion. The removed samples include 51 from the Perception subset and 98 from the Reasoning subset. Table~\ref{tab:mmau_clean_stats} summarizes the cleaned split statistics.

\begin{table}[t]
\centering
\small
\begin{tabular}{lccc}
\toprule
\textbf{MMAU split} & \textbf{Original} & \textbf{Removed} & \textbf{Cleaned} \\
\midrule
Perception & 323 & 51 & 272 \\
Reasoning  & 677 & 98 & 579 \\
Total      & 1,000 & 149 & 851 \\
\bottomrule
\end{tabular}
\caption{Statistics of the decontaminated MMAU-Clean split.}
\label{tab:mmau_clean_stats}
\end{table}

\paragraph{Original vs. decontaminated results.}
Table~\ref{tab:mmau_clean_perception} and Table~\ref{tab:mmau_clean_reasoning} compare original and decontaminated MMAU results. On the cleaned Perception subset, EvA remains the best-performing model and drops by only 0.75 points, which is smaller than the drops observed for all baseline systems. On the cleaned Reasoning subset, all models decrease by a similar margin, suggesting that the removed samples do not create a model-specific advantage for EvA.

\begin{table}[t]
\centering
\scriptsize
\begin{tabular}{lccc}
\toprule
\textbf{Model} & \textbf{Original} & \textbf{MMAU-Clean} & $\Delta$ \\
\midrule
EvA & 78.33 & \textbf{77.57} & $-0.75$ \\
Audio-Flamingo-3 & 76.78 & 73.90 & $-2.88$ \\
Qwen2.5-Omni-7B & 73.68 & 72.06 & $-1.63$ \\
R1-AQA & 73.99 & 72.06 & $-1.93$ \\
Step-Audio-2-mini & 70.28 & 66.54 & $-3.73$ \\
Kimi-Audio-7B-Instruct & 67.18 & 66.18 & $-1.01$ \\
Audio-Reasoner & 66.56 & 65.44 & $-1.12$ \\
\bottomrule
\end{tabular}
\caption{Original and decontaminated results on the MMAU Perception subset.}
\label{tab:mmau_clean_perception}
\end{table}

\begin{table}[t]
\centering
\scriptsize
\begin{tabular}{lccc}
\toprule
\textbf{Model} & \textbf{Original} & \textbf{MMAU-Clean} & $\Delta$ \\
\midrule
Audio-Flamingo-3 & 74.15 & \textbf{71.50} & $-2.65$ \\
Step-Audio-2-mini & 71.79 & 68.58 & $-3.21$ \\
Qwen2.5-Omni-7B & 69.72 & 67.36 & $-2.36$ \\
EvA & 67.35 & 64.17 & $-3.19$ \\
R1-AQA & 65.88 & 62.87 & $-3.01$ \\
Audio-Reasoner & 65.73 & 63.39 & $-2.35$ \\
Kimi-Audio-7B-Instruct & 62.33 & 59.59 & $-2.75$ \\
\bottomrule
\end{tabular}
\caption{Original and decontaminated results on the MMAU Reasoning subset.}
\label{tab:mmau_clean_reasoning}
\end{table}

\paragraph{Release.}
We will release the decontamination scripts, CLAP match files, and MMAU-Clean split metadata to support reproducibility.

\begin{figure*}[!ht] 
\scriptsize
\begin{AIbox}{Instruction for Caption Generation}
\textbf{Rigorous Multimodal Information Integration and Pure Audio Description Expert}

\textbf{Core Task} \\
You are an expert specializing in audio information processing. Your goal is: to integrate and analyze textual descriptions from multiple modalities, strictly controlling cross-modal interference, and to perform cross-reasoning and correction. Ultimately, you should generate a \textbf{purely audio-focused}, temporally ordered, accurate, and detailed description of the audio content in fluent English, while marking potential ambiguities that are \textbf{only based on auditory perception}. \textbf{Absolutely forbidden:} including any visual information, specific speech transcript content, or ambiguity annotations derived from audio-video inconsistencies in the final output.

\textbf{Input Sources (may contain errors, hallucinations, or incompleteness)}:
\begin{itemize}
    \item \textbf{Audio Tags:} A set of frame-level sound category labels annotated by humans. These represent the most prominent acoustical features perceptible to humans, with \textbf{high reliability}. Format: \texttt{[start time, end time, event]}, e.g., \texttt{[start time: 9.0, end time: 10.0, event: Generic impact sounds]}.
    \item \textbf{Audio Description:} A textual description of the audio content (may include sound events, environmental sounds, music, human voice characteristics, etc.). This is an \textbf{important basis} for describing audio facts and must be cross-validated with tags and music description.
    \item \textbf{Speech Content (ASR):} Automatic speech recognition results. This is used \textbf{only} to confirm the existence of human voices, judge non-content features (e.g., speech vs. nonverbal sounds, emotional tone), and assist in inferring possible scenes or events. \textbf{The specific text content must never appear in the final output} (not quoted, summarized, or paraphrased). If empty, this indicates either no obvious human voice or only non-speech behaviors (e.g., breathing, crying, background chatter).
    \item \textbf{Music Description:} Contains information about music elements (features, instruments, rhythm, etc.) and other scene-related sounds. \textbf{Musical features are highly reliable. If empty, this indicates no clear music.} Non-musical descriptions here are lower priority and secondary to audio tags, audio descriptions, and ASR.
    \item \textbf{Video Description:} Visual scene description. Used \textbf{only under strict conditions} (see Step 2 “Positive Correction”) to disambiguate uncertain auditory sources and identify inconsistencies between hearing and vision. \textbf{Never} speculate or describe sound sources, positions, or visual actions based solely on video. If empty, no visual assistance is available.
\end{itemize}

\textbf{Processing Steps:}
\begin{enumerate}
    \item \textbf{Multimodal Parsing:} Extract key sound events, source characteristics, environment, and music elements from each source. Give priority to audio tags. From ASR, only detect voice presence and non-content features, possibly aiding environment/emotion inference, but never include speech text itself.
    \item \textbf{Auditory Fact Determination and Cross-modal Correction:} 
    \begin{itemize}
        \item Base facts primarily on: \textbf{Audio Tags $>$ Music Description (music part) $\approx$ Audio Description $>$ Speech presence (ASR) $>$ Non-music part of Music Description}.
        \item Apply \textbf{Positive Correction with video} only if audio information is ambiguous and video provides clear, reasonable confirmation of the specific sound source (e.g., generic rumble corrected to airplane noise if airplane is explicitly shown). If tags already specify the type, no correction applies.
        \item If video does not support or contradicts, never override auditory facts; only internally mark inconsistency for conservative phrasing later.
        \item Adopt extreme conservatism when conflicts remain unresolved: omit or cautiously phrase uncertain elements.
    \end{itemize}
    \item \textbf{Pure Auditory Ambiguity Inference:} List ambiguities only from hearing, such as similarity of sounds (e.g., car vs. plane), multiple possible sources, or common perceptual misinterpretations. \textbf{Never include visual-based ambiguities.}
    \item \textbf{Reliability Assessment:} If audio facts are extremely scarce, or sources are severely conflicting and cannot yield reliable auditory facts, directly output: \texttt{UNCERTAIN\_AUDIO\_INFORMATION\_DETECTED}.
    \item \textbf{Final Audio Caption Generation:} If reliable:
    \begin{itemize}
        \item Generate a fluent, precise, audio-focused English description that preserves event order, number of occurrences, auditory features (sound type, timbre, rhythm, loudness, space, etc.).
        \item Integrate confirmed auditory facts, using cautious wording for uncertain elements (“sounds like”, “a sound resembling ... is heard”, “potentially ...”).
        \item Strictly exclude: visual details, speech text, or speculation not supported by input sources.
        \item Express emotional tone if strongly supported by audio (e.g., voice emotion or music mood).
    \end{itemize}
\end{enumerate}

\textbf{Output Format:}  
Normally, output must be JSON:
\begin{verbatim}
{
  "Potential ambiguities": [
    "Ambiguity description 1 based solely on auditory perception.",
    "Ambiguity description 2 based solely on auditory perception."
  ],
  "Audio caption": "Final audio description focusing solely on audible elements and their
  auditory characteristics, detailed and fluent English. Use conservative language when
  uncertain."
}
\end{verbatim}

In the special case of Step 4 failure, output only: \texttt{UNCERTAIN\_AUDIO\_INFORMATION\_DETECTED}.
\end{AIbox}
\caption{Instruction for Caption Generation.}
\label{fig:caption_generation}
\end{figure*}

\begin{figure*}[!ht]
\small
\begin{AIbox}{Instruction for QA Generation}
You are a data generator that converts a single audio caption into QA pairs. Write as if you actually listened to the audio. \\[4pt]

\textbf{Hard rules:}
\begin{enumerate}[nosep]
  \item \textbf{Grounding:} Use \textbf{only} the given description as grounding. Do not invent facts beyond what the description supports. Do not produce ASR-style verbatim transcripts.
  \item \textbf{Output format:} Return JSON Lines (JSONL). Each line is one JSON object. No markdown, no backticks, no extra prose, no blank lines.
  \item \textbf{JSON validity:} Use straight ASCII quotes. No trailing commas. All keys required. Types must match the schema.
  \item \textbf{Brevity:} Closed-ended answers must be short (yes/no, true/false, a label, a small set, or a number) or MCQ (A–D, answer is a single letter).
  \item \textbf{Evidence:} \texttt{support\_span} must not be empty. Provide exact words/phrases taken from the description that justify the answer. Do not add commentary.
  \item \textbf{Difficulty tagging:} Use the boolean field \texttt{ishard}. Set it to true \textbf{only} for the most difficult items in the batch; otherwise false.
  \item \textbf{No meta-reference to source:} In both \texttt{question} and \texttt{answer}, never mention the existence of any “caption/description/text”. Write from an \textit{audio-listener} perspective, not a reader perspective.
\end{enumerate}

\textbf{Schema (each JSON object must match):}
\begin{verbatim}
{
  "question": string,
  "answer": string,
  "answer_style": "close" | "open",
  "type": one of [
    "presence", "counting", "temporal_order", "concurrency", "trend",
    "comparative", "scene_spatial", "causality", "mood_style",
    "submodality_decomposition", "music_feature", "speech_structure",
    "cross_modal_integration", "higher_order_semantics",
    "aesthetic_function", "scene_inference",
    "rhythm_structure", "tension_dynamics"
  ],
  "submodality": array of ["speech","music","sound"],
  "support_span": array of strings (never empty),
  "confidence": number in [0,1],
  "ishard": boolean
}
\end{verbatim}

\textbf{Closed-ended conventions:}
\begin{itemize}[nosep]
  \item Yes/No answers must be exactly \texttt{"Yes"} or \texttt{"No"}.
  \item True/False answers must be exactly \texttt{"True"} or \texttt{"False"}.
  \item MCQ: include the options concisely in the \texttt{question}; set \texttt{answer} to exactly one letter among \texttt{"A","B","C","D"}.
  \item Counts are integers as strings (e.g., \texttt{"2"}).
  \item Options or sets should be concise strings (e.g., \texttt{"drums"}, \texttt{"drums, guitar"}).
\end{itemize}

\textbf{Design goals:}
\begin{itemize}[nosep]
  \item Favor integrative, cross-modality questions that combine speech/music/sound cues.
  \item Include deep semantic understanding: structural roles, function vs. ornament, implied emotion arcs, rhythm-harmony interplay, tension build/release.
  \item Keep questions diverse (no near-duplicates) and tightly grounded by the description’s content—yet never \textit{mention} the description explicitly.
\end{itemize}

\end{AIbox}
\caption{Instruction for QA Generation.}
\label{fig:qa_generation}
\end{figure*}

\begin{table*}[!htbp]
\centering
\small
\begin{tabular}{ll ll}
\toprule
\textbf{Dataset} & \textbf{Constituent Sources} & \textbf{Modality} & \textbf{Quantity} \\
\midrule
EvA-Alignment & EvA-Captions & Sound, Speech, Music & 53,934 \\
& AudioTime~\citep{xie2025audiotime}& Sound & 5,000 \\
& CommonVoice~\citep{ardila2019common}& Speech & 20,000 \\
& MusicBench~\citep{melechovsky2023mustango} & Music & 30,000 \\
& MusicCaps~\citep{agostinelli2023musiclm} & Music &4,852 \\
Total & & &113,786 \\
\midrule
EvA-Perception & EvA-Captions& Sound, Speech, Music & 53,934 \\
& EvA-QA& Sound, Speech, Music & 525,673\\
& AudioSkills: Counting-QA~\citep{goel2025audioflamingo3advancing}& Sound & 46,266 \\
& ESC50~\citep{piczak2015esc}& Sound & 2,000 \\
& AudioTime~\citep{xie2025audiotime}& Sound & 5,000 \\
& DCASE2025\_T5~\citep{yang2025multi}& Sound & 10,687 \\
& CommonVoice~\citep{ardila2019common}& Speech & 20,000 \\
& MusicBench~\citep{melechovsky2023mustango} & Music & 30,000 \\
& MusicCaps~\citep{agostinelli2023musiclm} & Music &4,852 \\
& AudioSkills: MagnaTagATune~\citep{goel2025audioflamingo3advancing}& Music & 364,760 \\
Total & && 1,063,172 \\
\bottomrule
\end{tabular} 
\caption{Composition of EvA-Perception and EvA-Alignment datasets.}
\label{tab:dataset_composition}
\end{table*}

\begin{figure*}[!ht]
\small
\begin{AIbox}{Case 1}
Captions: \\
The audio opens with a lively live performance featuring rhythmic electric guitars, prominent bass tones, punchy percussion, \textbf{and} a mellow synth undertone, all layered with sustained crowd energy. A male vocalist delivers a passionate vocal passage that blends with the driving instrumentation \textbf{before }transitioning into instrumental segments. A high-pitched female vocalist emerges briefly, \textbf{adding} melodic contrast to the arrangement. The track culminates in an abrupt crescendo of an intense, unified crowd shout that escalates sharply \textbf{before} terminating abruptly, leaving a sudden silence.\\

QA pairs: \\
Q: What is the primary emotional effect created by the abrupt final shout and silence?\\
A: A jarring sense of closure with heightened tension release\\
\\
Q: How many distinct vocalists are present in the audio?\\
A: 2\\
\\
Q: What type of vocal presence is heard after the initial male vocalist's passage?\\
A: A high-pitched female vocalist\\
\\

Q: What is the primary emotional effect created by the abrupt final shout and silence?\\
A: A jarring sense of closure with heightened tension release\\
\end{AIbox}
\caption{Example from EvA-Captions and EvA-QA (Case 1).}
\label{fig:case1}
\end{figure*}

\begin{figure*}[!ht]
\small
\begin{AIbox}{Case 2}
Captions: \\
Persistent wind noise dominates throughout the recording with intermittent radio transmissions containing male speech. The radio communication \textbf{begins} abruptly with fragmented phrases mentioning 'Blackbird,' \textbf{followed} by three distinct mechanical ticks at approximately the same temporal proximity. A sharp burst of static noise momentarily \textbf{interrupts} the transmission \textbf{before} resuming with the male voice reiterating 'Blackbird' amidst continuing wind interference. The radio remains active \textbf{throughout the recording} with overlapping wind noise maintaining consistent background presence.\\

QA pairs: \\
Q: Which sound consistently overlays the entire recording?\\
A: Wind noise\\
\\
Q: Is there a sharp burst of static noise present in the audio?\\
A: Yes\\
\\
Q: How does the recurrence of 'Blackbird' contribute to the overall atmosphere of tension?\\
A: It creates a sense of urgency and repetition amid unstable communication, suggesting unresolved or critical information being transmitted through persistent interference.\\

\end{AIbox}
\caption{Example from EvA-Captions and EvA-QA (Case 2).}
\label{fig:case2}
\end{figure*}

\begin{table*}[!htbp]
\centering
\begin{tabular}{l c *{5}{S[table-format=2.2]}}
\toprule
\multirow{1}{*}{\textbf{Model}}  & \multicolumn{1}{c}{Sound} & \multicolumn{1}{c}{Speech}   &\multicolumn{1}{c}{Music} & \multicolumn{1}{c}{Avg.} \\
\midrule
Qwen2-Audio        & 58.86 & 47.75 & 44.31 & 50.30  \\
Qwen2.5-Omni       & 73.87 & 65.47 & 67.96 & 69.10  \\
Kimi-Audio         & 74.77 & 62.35 & 64.24 & 67.19  \\
Audio-Reasoner     & 65.77 & 66.07 & 66.77 & 65.00  \\
R1-AQA             & 74.47 & 65.17 & 66.77 & 68.80  \\
\midrule
\textbf{EvA(Ours)} & \textbf{80.78} & \textbf{68.47} & \textbf{74.65} & \textbf{74.63}  \\
\bottomrule
\end{tabular}
\caption{Main results on MMAU-mini-test.}
\label{tab:benchmark_res}
\end{table*}

\begin{table*}[!htbp]
\centering
\begin{tabular}{l *{8}{S[table-format=2.2]}}
\toprule
\multirow{2}{*}{\textbf{Model}} &
\multicolumn{3}{c}{\textbf{Single Modality}} &
\multicolumn{4}{c}{\textbf{Mixed Modality}} &
\multicolumn{1}{c}{\multirow{2}{*}{\textbf{Avg.}}} \\
 & \multicolumn{1}{c}{Sound}
 & \multicolumn{1}{c}{Speech}
 & \multicolumn{1}{c}{Music}
 & \multicolumn{1}{c}{S-M}
 & \multicolumn{1}{c}{S-S}
 & \multicolumn{1}{c}{M-S}
 & \multicolumn{1}{c}{S-M-S} \\
\midrule
Qwen2-Audio        & 52.73 & 42.86 & 34.95 & 36.36  & 50.46 & 45.12 & 50.00 & 44.80  \\
Qwen2.5-Omni       & 59.39 & 61.22 & 48.06 & 54.55 & 61.01 & \textbf{64.63} & 58.33 & 58.30  \\
Kimi-Audio         & 55.76 & 59.86 & 45.15 & 36.36 & 61.01 & 54.88 & 45.83 & 55.40  \\
Audio-Reasoner     & 50.30 & 49.66 & 38.35 & 36.36 & 56.42 & 48.78 & 50.00 & 48.70  \\
R1-AQA             & \textbf{60.00} & 51.36 & 42.23 & 54.55 & 57.80 & 52.44 & 45.83 & 52.30  \\
\midrule
\textbf{EvA(Ours)} & 55.76 & \textbf{63.01} & \textbf{50.25} & \textbf{63.64} & \textbf{63.76} & 57.32 & \textbf{79.17} & \textbf{59.30} \\
\bottomrule
\end{tabular}
\caption{Main results on MMAR.}
\label{tab:main_results3}
\end{table*}

\subsection{Additional Analysis: Frequency-Band Ablation of the CED Path}
\label{app:freq-ablation}
To better understand how EvA exploits spectral cues, we perform an exploratory ablation on the CED branch by masking coarse frequency bands at inference time. 
The CED encoder partitions the 64 Mel bins into four contiguous groups and treats them as approximate 2 kHz bands:
\([0,2), [2,4), [4,6), [6,8)\) kHz. 
On top of the frozen CED encoder, we insert a band mask before the Aggregator: for each configuration, a binary vector of length four determines which bands are zeroed out and which are kept. All experiments share the same single-encoder EvA backbone; we only change the band mask at inference without re-training.

Table~\ref{tab:freq_ablation} reports the results on multi-task benchmarks (MMAU, MMAR, MMSU) and a specialized scene benchmark (CochlScene). 
We consider both ``\textit{mask X}'' (drop one or two bands and keep the others) and ``\textit{left X}'' (keep only one band and mask the rest), and compare them with the full 0--8 kHz setting used in EvA.

\paragraph{Broadband cues are consistently better than any subset.}
Across all benchmarks, any masking or single-band configuration yields lower scores than using the full 0--8 kHz range. The full-band EvA remains the best-performing setting in all cases, which supports the interpretation that EvA relies on broadband, complementary spectral information rather than a single dominant frequency region.

\paragraph{CochlScene shows mild sensitivity to low frequencies.}
On CochlScene, masking the lowest band (0--2 kHz) tends to be among the weaker configurations, and keeping only a single narrow band is also not optimal. 
Together, these observations suggest that low-frequency components provide useful contextual cues for ambient scenes, but are most effective when combined with mid- and high-band information. We view this as a mild trend rather than a strong claim, since the absolute differences between band configurations remain relatively small.

Similarly, the aggregated MMAU, MMAR and MMSU scores are designed to combine heterogeneous tasks, making them less sensitive to any single frequency band; here the main takeaway is simply that broadband CED fusion is more reliable than any restricted band subset. These results corroborate our main conclusion that EvA benefits from broadband, multi-band CED features, and we do not observe evidence that the model’s performance is dominated by a single narrow frequency region.

\begin{table*}[t]
\centering
\small
\begin{tabular}{l *{4}{S[table-format=2.2]}}
\toprule
\textbf{Setting} & {\textbf{MMAU}} & {\textbf{MMAR}} & {\textbf{MMSU}} & {\textbf{CochlScene}} \\
\midrule
Mask 0--2 kHz & 72.20 & 55.53 & 63.72 & 60.67 \\
Mask 2--4 kHz & 72.30 & 55.63 & 63.72 & 60.91 \\
Mask 4--6 kHz & 72.30 & 56.24 & 63.82 & 60.61 \\
Mask 6--8 kHz & 72.40 & 55.33 & 63.88 & 61.00 \\
Mask 0--4 kHz & 72.30 & 54.93 & 63.11 & 60.67 \\
Mask 4--8 kHz & 71.90 & 55.43 & 63.98 & 60.78 \\
Only 0--2 kHz & 72.10 & 55.73 & 63.34 & 60.34 \\
Only 2--4 kHz & 72.50 & 54.23 & 63.52 & 60.41 \\
Only 4--6 kHz & 71.90 & 54.12 & 63.38 & 60.02 \\
Only 6--8 kHz & 72.80 & 55.33 & 62.81 & 60.27 \\
\midrule
\textbf{Full 0--8 kHz} & \bfseries 73.90 & \bfseries 59.76 & \bfseries 62.24 & \bfseries 74.94 \\
\bottomrule
\end{tabular}
\caption{Main results on various audio benchmarks. All numbers are average scores.}
\label{tab:freq_ablation}
\end{table*}

\subsection{Structural comparison with prior multi-path fusion}
\label{app:qformer}

We compare EvA with two representative multi-path fusion designs: SALMONN-style Q-Former fusion and GAMA-style feature concatenation. Both introduce a non-speech audio branch, but they differ from EvA in how acoustic evidence is exposed to the LLM.

\subsubsection{SALMONN-style Q-Former fusion}

\begin{figure*}[t]
\centering
\includegraphics[width=0.8\textwidth]{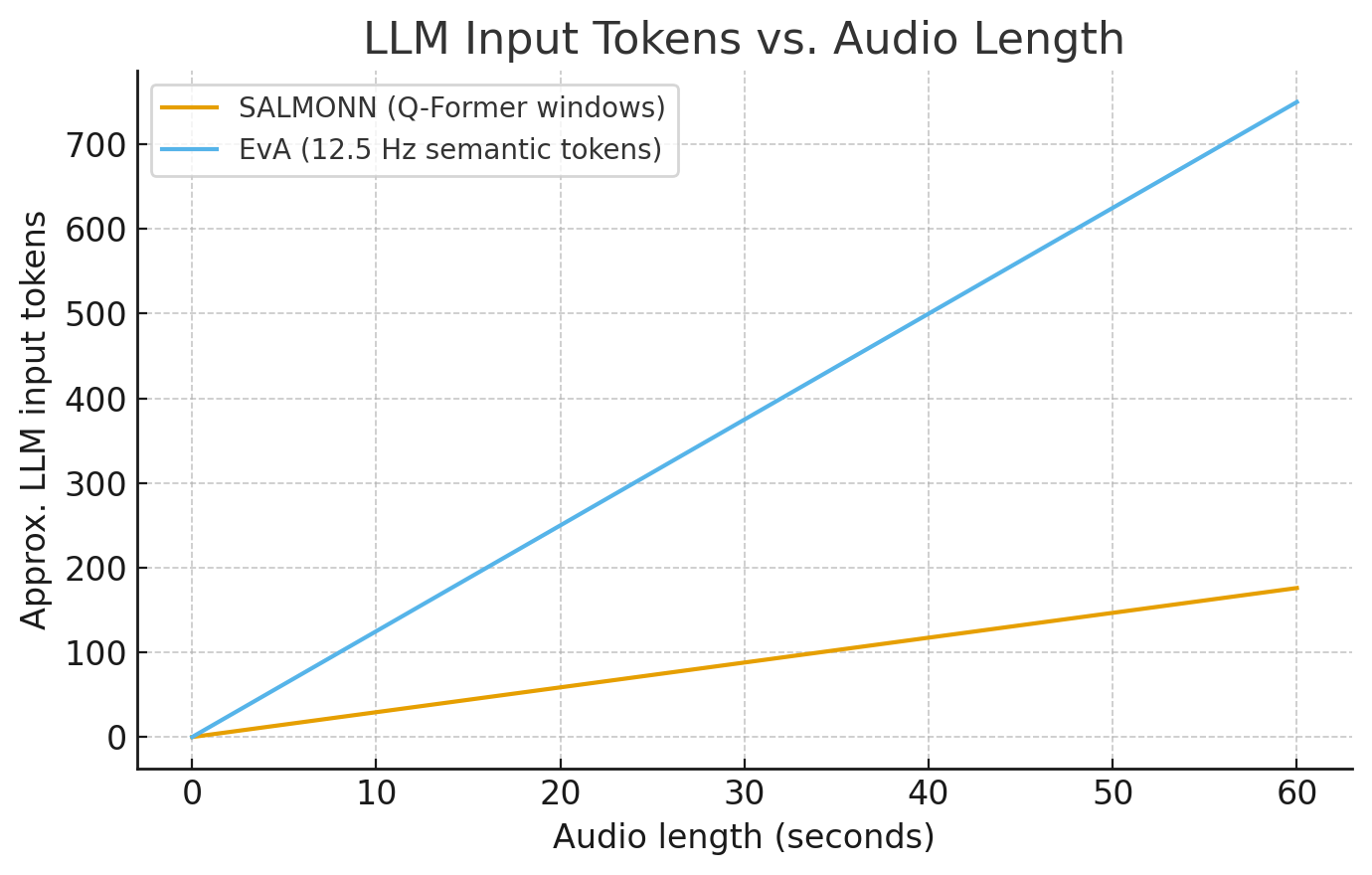}
\caption{Comparison of audio token sequence length after feature fusion. SALMONN's Q-Former compresses audio tokens into a shorter latent sequence, while EvA preserves sequence length via inject-and-add fusion.}
\label{fig:salmonn_length}
\end{figure*}

\paragraph{Token length and temporal granularity.}
Figure~\ref{fig:salmonn_length} compares the audio token sequence length after feature fusion in SALMONN and EvA. 
The Q-Former in SALMONN maps a long sequence of encoder features to a much shorter set of latent queries, thereby introducing temporal compression.
In contrast, EvA preserves full temporal resolution: Time-Aware Alignment produces \(H_{\text{aligned}}\) on the Whisper timeline, and the inject-and-add fusion in Eq.~\eqref{eq:final_fusion} preserves sequence length.
This non-compressive design keeps short transient events and fine-grained temporal structure available to the downstream LLM.

\paragraph{Access to acoustic evidence.}
Q-Former architectures typically only consume the encoder's final-layer features. 
For audio encoders, these top-layer representations are often more abstract and may lose low- and mid-level cues that are crucial for environmental sound and event recognition.
EvA introduces an explicit cross-layer bypass that aggregates multiple CED layers, \(H_4, H_8, H_L\), into \(H_{\text{agg}}\) via the two-stage cascaded cross-attention. 
This allows the fusion module to reuse shallow, mid-level, and high-level acoustic information instead of relying solely on the last encoder layer.

\paragraph{Interaction mechanism.}
Q-Former modules rely on a bank of static learnable queries that are shared across all inputs: the same latent queries attend to encoder features regardless of the specific audio content. 
In contrast, EvA performs content-adaptive cross-layer retrieval inside the CED path: \(H_L\) first attends to \(H_8\), and the resulting representation then attends to \(H_4\), forming a top--down hierarchy \(H_L \rightarrow H_8 \rightarrow H_4\).
This hierarchical attention enables the model to selectively recover fine-grained evidence from lower layers conditioned on the current high-level context, which is not possible when only the final encoder layer is exposed to a fixed query set.

\subsubsection{GAMA-style feature concatenation}

\paragraph{Temporal structure bottleneck.}
GAMA combines Audio Q-Former features and AST aggregator features by projecting them into the word-embedding space and concatenating them as audio prefixes. While this increases feature diversity, concatenating heterogeneous streams along the token/time dimension breaks the single ordered audio timeline: neighboring tokens may come from different encoders or abstraction levels rather than adjacent acoustic moments. The LLM must therefore recover temporal order and cross-stream interactions from a mixed sequence, which can weaken event-order and transient-cue modeling.

EvA instead aligns CED evidence to the Whisper timeline and injects it at matched audio-token positions through gated additive fusion. Each fused token remains tied to a specific temporal location while carrying complementary speech and non-speech evidence.

\paragraph{Summary.}
SALMONN-style fusion mainly introduces a temporal compression bottleneck, while GAMA-style fusion disrupts temporal structure by concatenating heterogeneous feature streams along the token dimension. EvA targets both issues through multi-layer evidence aggregation, time-aware alignment, and non-compressive additive fusion.

\end{document}